\documentclass[grl]{agutex}

\usepackage{xcolor}
\usepackage{graphicx}
\usepackage{lscape}
\usepackage{rotating}
\authorrunninghead{MERCIER ET AL.}

\titlerunninghead{LABORATORY MODELING OF M$_2$ INTERNAL TIDE AT LUZON STRAIT}

\authoraddr{Corresponding author: M. Mercier, MIT (3-351K), 77 Mass. Ave., Cambridge, MA 02139, USA. (mmercier@mit.edu)}

\begin{document}

%
%

\title{Large-scale, realistic laboratory modeling of M$_2$ internal tide generation at the Luzon Strait}
%
%

%
%



\authors{Matthieu J. Mercier,\altaffilmark{1} Louis Gostiaux,\altaffilmark{2}  Karl Helfrich,\altaffilmark{3} Joel Sommeria,\altaffilmark{4} Samuel Viboud,\altaffilmark{3} Henri Didelle,\altaffilmark{3}   Sasan J. Ghaemsaidi,\altaffilmark{1} Thierry Dauxois,\altaffilmark{5} and Thomas Peacock\altaffilmark{1}}
\altaffiltext{1}{Department of Mechanical Engineering, Massachusetts Institute of Technology, Cambridge, MA, USA.}
\altaffiltext{2}{Laboratoire de M\'ecanique des Fluides et dÕAcoustique, UMR 5509, CNRS, Universit\'e de Lyon, \'Ecully, France.}
\altaffiltext{3}{Woods Hole Oceanographic Institution, Woods Hole, MA, USA.}
\altaffiltext{4}{CNRS/Grenoble-INP/UJF-Grenoble1, LEGI UMR 5519, Grenoble, France.}
\altaffiltext{5}{Universit\'{e} de Lyon, Laboratoire de Physique, \'{E}cole Normale Sup\'{e}rieure de Lyon, CNRS, Lyon, France.}


%
%


\begin{abstract}

The complex double-ridge system in the Luzon Strait in the South China Sea (SCS) is one of the strongest sources of internal tides in the oceans, associated with which are some of the largest amplitude internal solitary waves on record.
An issue of debate, however, has been the specific nature of their generation mechanism. To provide insight, we present the results of a large-scale laboratory experiment performed at the Coriolis platform. The experiment was carefully designed so that the relevant dimensionless parameters, which include the excursion parameter, criticality, Rossby and Froude numbers, closely matched the ocean scenario.
The results advocate that a broad and coherent weakly-nonlinear, three-dimensional, M$_2$ internal tide that is shaped by the overall geometry of the double-ridge system is radiated into the South China Sea and subsequently steepens, as opposed to being generated by a particular feature or localized region within the ridge system.
\end{abstract}

%
%

%

\begin{article}

%
%
\section{Introduction}

The Luzon Strait (LS), a complex double-ridge system between Taiwan and the Philippines, generates some of the strongest internal tides found anywhere in the oceans \citep{bib:Alford2011}. Consequently, a striking characteristic of the South China Sea (SCS) to the west of the LS is the regular observation of large amplitude Internal Solitary Waves (ISWs) with amplitudes up to 170\,m and speeds of 3\,m/s, taking about 33\,hours to traverse the SCS basin \citep{bib:Alford2010}. A summary of observations over a three-year time period reveals that these ISWs are primarily directed slightly north of west, with surface signatures of the ISWs are rarely observed east of 120.3$^{\circ}$ \citep{bib:Jackson2009}.

In recent years, there has been a concerted effort to more clearly establish the early stages of the formation of ISWs in the SCS.
The lee wave mechanism \citep{bib:Maxworthy1979}, initially alluded to as a candidate process \citep{bib:Ebbesmeyer1991}, is now considered inappropriate \citep{bib:Alford2010} due to the lack of reported detection between the ridges, or observations in satellite imagery, of ISWs east of $120.3^o$ \citep{bib:Jackson2009}. Instead, it is now widely accepted that a westward propagating, weakly-nonlinear, mode-1 internal tide generated at the LS steepens to form ISWs \citep{bib:Lien2005,bib:Li2011}.

A prevailing theme has been to identify generation hot spots within the LS. Two-dimensional models \citep{bib:Zhao2006,bib:Alford2010} and numerical simulations employing idealized or actual transects containing purported generation sites \citep{bib:Buijsman2010a,bib:WarnVarnas2010}, have been used in this regard, producing discordant results. The inconsistency of the outcome of two-dimensional studies is most likely due to the inherently three-dimensional nature of the internal tide generation process, with extended sections of the two ridges playing a key role $\!$\citep{bib:Zhang2011,bib:Guo2011}.

Here, we present the results of a large-scale, realistic, laboratory project of internal tide generation in the LS. The experiments, performed at the Coriolis Platform of LEGI (Grenoble, France), model key constituents of the generation process: three-dimensional bathymetry, nonlinear stratification, semi-diurnal (M$_2$) tidal forcing, and background rotation. The goal of the experiments is to elucidate the geometry and characterize the nonlinearity of the radiated three-dimensional M$_2$ internal tide, thereby providing insight into the consequent generation of ISWs in the SCS.
Due to the difference in the dissipation processes, laboratory experiments are a valuable complement to realistic three-dimensional numerical simulations. 
Any result reproduced by both approaches can be considered robust.
We describe the experimental arrangement in \S2 and discuss the scalings chosen to preserve dynamic similarity with the ocean in \S 3. Results for the M$_2$ barotropic and baroclinic (internal) tides are presented in \S 4 and \S 5 respectively. The nonlinearity of the internal tide is characterized in \S 6 before discussion and conclusions in \S 7.

\section{Experimental configuration}

\begin{figure*}
\centering
\includegraphics[width=0.95\linewidth]{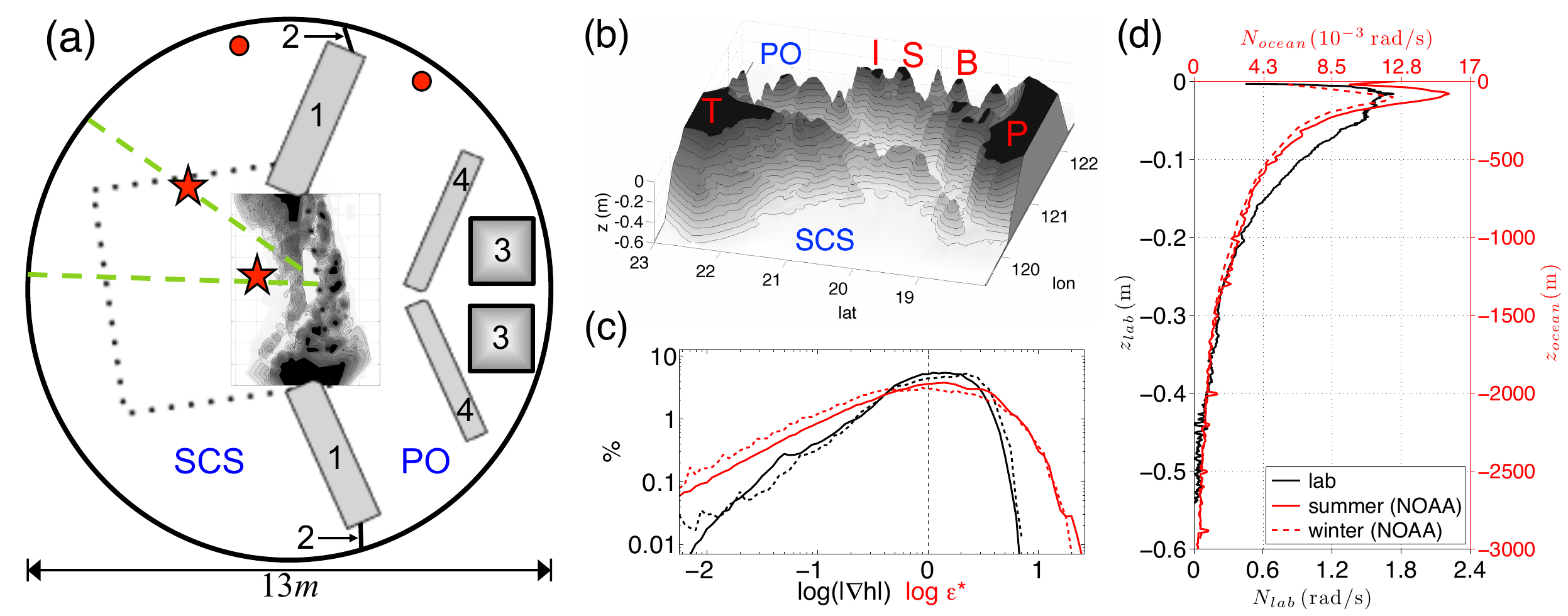}
\caption{(a) Schematic of the experiment. Vertical partitions (1) extend from Taiwan and The Philippines to the side of the tank . After filling, the PO and SCS are separated using inserted barriers (2) and the barotropic tide is generated using prismatic plungers (3) located behind vertical walls (4). Dashed green lines and the dotted square indicate the vertical PIV planes and overhead field of view, respectively. Red stars and circles indicate the location of the CT and acoustic probes, respectively.
(b) The experimental topography, with Taiwan (T), The Philippines (P), Itbayat Island (I), Sabtang \& Batan Islands (S) and Babuyan Island~(B) indicated. (c) Distributions of the topographic slope (black) and the criticality parameter $\varepsilon^*$ (red) for the experiments (solid) and the ocean (dashed). The ocean slopes have been multiplied by $20$ to match the laboratory range. (d) Typical winter and summer stratifications (red) in the LS, compared to a laboratory stratification (black).}\label{fig1}
\end{figure*}

A schematic of the experiment is presented in Fig.\,\ref{fig1}(a). It was performed at the Coriolis Platform, the world's largest turntable facility for geophysical fluid dynamics experiments, in order to incorporate the 
appropriate influence of background rotation and to curtail viscous effects.
A 5\,m long, 2\,m wide and 0.6\,m tall scale model of the LS constructed based on topographic data provided by the Ocean Data Bank, National Science Council, Taiwan, was positioned near the center of the tank. Partition barriers were introduced north and south of the model, representing Taiwan and the Philippines, respectively. Damping material was affixed to the inner walls of the tank to damp reflections effectively, as discussed in \S \ref{sec:BC}.
One of a pair of vertically oscillating prismatic plungers was used to generate the equivalent of a M$_2$ barotropic tide~\citep{bib:Biesel1951}. Located in the Pacific Ocean (PO) basin behind vertical partitions blocking internal waves emanating from the plunger itself, each plunger was tested individually. No differences were found in the resulting tidal flows.

A three-dimensional rendering of the experimental topography is presented in Fig.\,\ref{fig1}(b). The horizontal and vertical scalings of the topography compared to the ocean were 1:100,000 and 1:5,000, respectively. To ensure these choices did not affect the dynamical processes being modeled, other physical quantities such as the stratification, tidal forcing and background rotation frequencies were appropriately rescaled, as described in \S 3. Only topographic features above $3000$ m deep were reproduced, since the weak deep-ocean stratification has negligible effect on the wavelengths and phase speeds of the vertical modes. Several metrics were calculated to ascertain the quality of reproduction of the original ocean data set; one measure, the cross-correlation coefficient, was $0.90\pm0.03$, indicating a good transcription from the ocean to the experiment. Over the same domain, the distribution of slopes, ${\bf\nabla} h$, presented in Fig.\,\ref{fig1}(c), confirms that the laboratory topography was indeed 20 times steeper than that of the ocean and with the same distribution.

A nonlinear density stratification was established using salt water introduced via a computer-controlled double-bucket system (see Fig.\,\ref{fig1}(d)). Velocity field data was obtained using Particle Image Velocimetry (PIV). For horizontal velocity field measurements, small particles of diameter 350 $\mu$m with a density in a narrow range ($\pm 1$ kg/m$^3$) corresponding to the base of the pycnocline were used to seed the fluid, ensuring they would clearly reveal horizontal motions associated with mode-1 and mode-2 internal tides; the particles were illuminated from below by submerged white lights. For vertical velocity field measurements, a moveable laser sheet in a vertical plane cutting across the ridge illuminated similar particles distributed uniformly in density to cover the entire fluid depth. Local density oscillations in the radiated wave field were measured using conductivity and temperature (CT) probes, and acoustic probes on either side of the ridge measured the surface deflection due to the barotropic tide with an accuracy of 0.05\,mm.


\section{Governing parameters}

\begin{table*}
\centering
\begin{tabular}{lllclccccccc}
\hline
& $\omega$ (rad/s) & $f$ (rad/s) &  $N_m$ (rad/s) & $N_b$ (rad/s) &  $U_0$ (m/s) & $\delta$ (m) & $H$ (m) & $L$ (m) & $h_0$ (m) & $\nu$ (m$^2$/s)  \\
\hline
{\it Luzon} & $1.40\!\times\!10^{-4}$ & $5.00\!\times\!10^{-5}$ &  $1.57\!\times\!10^{-2}$ & $3.65\!\times\!10^{-4}$ &  $0.1\quad$ & 100 & $3.0\!\times\!10^3$ & $10^{5}$ & $1.5\!\times\!10^{3}$ & $10^{-4}$ \\
{\it Batanes} & $1.40\!\times\!10^{-4}$ & $5.00\!\times\!10^{-5}$ &  $1.57\!\times\!10^{-2}$ & $1.28\!\times\!10^{-3}$ &  $1.0\quad$ & 100 & $1.5\!\times\!10^3$ & $10^{4}$ & $8.0\!\times\!10^{2}$ & $10^{-4}$ \\
{\it Lab$_L$} & $3.86\!\times\!10^{-1}$ & $1.38\!\times\!10^{-1}$ & $2.21\quad$ & $5.15\!\times\!10^{-2}$ & $2.76\!\times\!10^{-3}$ & 0.02 & $0.6\quad$ & $1.0\quad$ & $0.30\quad$ & $10^{-6}$ \\
{\it Lab$_B$} & $3.86\!\times\!10^{-1}$ & $1.38\!\times\!10^{-1}$ & $2.21\quad$ & $1.80\!\times\!10^{-1}$ & $2.76\!\times\!10^{-2}$ & 0.02 & $0.3\quad$ & $0.1\quad$ & $0.16\quad$ & $10^{-6}$ \\
{\it $\frac{\textrm{Lab}_L}{\textrm{Luzon}}$} & $2.76\!\times\!10^{3}$ & $2.76\!\times\!10^{3}$ & $1.41 \!\times\!10^2$ & $1.41 \!\times\!10^2$ & $2.76\!\times\!10^{-2}$ & $2\!\times\!10^{-4}$ & $2\!\times\!10^{-4}$ & $10^{-5}$ & $5\!\times\!10^{3}$ & $10^{2}$ \\
\hline
\end{tabular}
\caption{Characteristic values of key dimensional parameters for internal tide generation on the scale of the Luzon Strait ({\it Luzon}) and on the local scale ({\it Batanes}). The corresponding laboratory values are {\it Lab$_L$} and {\it Lab$_B$}, respectively. The scaling factor relating laboratory values to the ocean ones is indicated at the last line.}
\label{tab1}
\end{table*}

\begin{table*}
\centering
\begin{tabular}{llccccccccccc}
\hline
 & $h^*$ & $\delta^*$ & $h_0/L$ & $N^*$ & $\varepsilon^*$ & $Re^*$ & $Ro^*$ & $A^*$ &$Lo^*$ & $Fr_1^*$ & $Fr_2^*$\\
\hline
 {\it Luzon} & $0.50$ & $3.3\!\times\!10^{-2}$ & $1.5\!\times\!10^{-2}$ & $43.0$ & $[0.004 - 14]$ & $1.0\!\times\!10^{8}$ & $2.0\!\times\!10^{-2}$ & $7.14\!\times\!10^{-3}$ & $2.5\!\times\!10^{-2}$ & $3.6\!\times\!10^{-2}$ & $7.3\!\times\!10^{-2}$\\
 {\it Batan} & $0.53$ & $6.6\!\times\!10^{-2}$ & $8.0\!\times\!10^{-2}$ & $12.3$ & $[0.004 - 14]$ & $1.0\!\times\!10^{8}$ & $2.0\quad$ & $7.14\!\times\!10^{-1}$ & $2.7\!\times\!10^{-1}$ & $4.2\!\times\!10^{-1}$ & $8.3\!\times\!10^{-1}$ \\
 {\it Lab$_L$} & $0.50$ & $2.7\!\times\!10^{-2}$ & $0.3$ & $42.0$ & $[0.01 - 14]$ & $2.76\!\times\!10^{3}$ & $2.0\!\times\!10^{-2}$ & $7.15\!\times\!10^{-3}$ & $1.2\!\times\!10^{-2}$ & $2.1\!\times\!10^{-2}$ & $4.3\!\times\!10^{-2}$ \\
 {\it Lab$_B$} & $0.53$ & $5.5\!\times\!10^{-2}$  & $1.6$ & $1.82$ & $[0.01 - 14]$ & $2.76\!\times\!10^{3}$ & $2.0\quad$ & $7.15\!\times\!10^{-1}$ & $3.5\!\times\!10^{-1}$ & $8.9\!\times\!10^{-1}$ & $1.8$ \\
{\it $\frac{\textrm{Lab}_L}{\textrm{Luzon}}$} & $1.0$ & $0.82$ & $20$ & $0.98$ & $O(1)$ & $2.76\!\times\!10^{-5}$& $1.0$ & $1.00$ & $0.5$ &$0.6$ &$0.6$ \\
\hline
\end{tabular}
\caption{Characteristic values of key dimensionless parameters for the ocean and the measured experimental values. The ratio of laboratory to ocean values actually achieved is given at the end of the table.}\label{tab2}
\end{table*}

There are ten dimensional parameters that characterize internal tide generation at the LS: the barotropic tidal frequency ($\omega$) and velocity ($U_0$), the Coriolis frequency ($f$), the maximum and bottom value of the ocean stratifications ($N_m$, $N_b$), the depth of the pycnocline ($\delta$) and of the ocean ($H$), the vertical ($h_0$) and across-ridge horizontal ($L$) scales of the topography, and the fluid viscosity ($\nu$), be it kinematic or turbulent.
With two dimensions, there are thus eight governing dimensionless parameters. The practical choice of these parameters is established \citep{bib:GarrettKunze07}, but prudence is necessary when estimating their values. In particular, one must differentiate between dynamics associated with global and local scale processes. We therefore consider both a global scale ({\it Luzon}) for the entire LS  and a local scale ({\it Batanes}) characterized by the channel between Batan and Itbayat islands, which has been touted as a hot spot for internal tide generation.

The eight governing dimensionless parameters are: (1,2) $h^*=h_0/H$ and $\delta^*=\delta/H$,  the ratio of the topographic height, $h_0$, and the pycnocline depth, $\delta$, to the depth of the surrounding ocean, $H$, respectively; $(3)\,\,$the aspect ratio of the topographic features $h_0/L$; $(4)\,\,$the ratio of the upper and deep ocean stratifications, $N^*=N_m/N_b$; $(5)\,\,$the criticality, $\varepsilon^*={\bf\nabla} h/\sqrt{(\omega^2-f^2)/(N^2-\omega^2)}$, which is the ratio of the local topographic slope to the internal wave ray slope; 
(6,7)$\,\,$the Reynolds and Rossby numbers, $Re^*=U_0L/\nu$ and $Ro^*=U_0/fL$, respectively, these being the ratios of inertia to viscous and Coriolis forces. Three alternative definitions of the wave nonlinearity parameter are (8): the tidal excursion parameter, $A^*=U_0/\omega L$, which is the ratio of the barotropic tidal excursion to the horizontal scale of the topography; the Long number, $Lo^*=U_0/N_mh_0$, quantifying the tendency for upstream blocking and flow separation due to the stratification; or the Froude number, $Fr_n^*=U_0/c^n_\phi$, the ratio of the flow speed to the propagation speed $c^n_\phi$ of mode-$n$, with $n=1$ being the lowest baroclinic mode. A comparison of the dimensional and dimensionless parameters for the ocean and the lab is given in Tables\,\ref{tab1} and \ref{tab2}, respectively.

The values in Table \ref{tab2} reveals that the LS can be characterized as a weakly nonlinear system on the large scale, with a relevant influence of background rotation and topographic blocking; and a more nonlinear one on the local scale, with a large excursion parameter promoting lee waves generation and internal wave energy trapping by supercritical flow.
The increase by a factor 20 in the aspect ratio and the wave ray slope (based on the rescaled stratification in Fig.\,1(d)), preserving criticality as can be seen when comparing the ocean and laboratory distributions of $\varepsilon^*$ in Fig.\,1(c), is irrelevant in the long wave limit (equivalent to the hydrostatic approximation), which is expected to be valid in the generation process; this is of importance for the subsequent wave steepening leading to ISWs, however, as discussed in \S 6, which therefore is not reproduced in the experiment.
Overall the values of the dimensionless parameters for the LS and the laboratory model are consistent, except for $Re^*$. As such, the nature of the oceanic internal tide generation process will be faithfully reproduced in the experiments.

\section{The M$_2$ barotropic tide}

Experiments were first performed using unstratified water to investigate solely the barotropic tide.
Horizontal views of the flow at $z=-0.12$m (corresponding to $-600$m in the ocean) are compared with TPXO\,7.1 predictions \citep{bib:Egbert2002} in Fig.\,2(a) and (b); the ellipses and colorbars are scaled according to the parameters listed in Table \ref{tab1} so that, in principle, they should appear the same. The orientations of the tidal ellipses are in very good agreement; in both cases, the ellipses are almost flat, indicating a weak influence of background rotation for most of the strait, except in very shallow regions where ellipses are quite large, which stress the importance of Coriolis effects in setting the amplitude of the tidal velocity near shallow features.
The spatial distributions of the tidal amplitude are also very similar, the intensification of the flow with decreasing depth being well reproduced. Intense magnification above $z=-0.12$m could not be visualized due to the shadows cast by the laser light (coming from the west) striking the topography.

Figure 2(c) presents a vertical transect of the barotropic tide taken along the blue dashed line in Figure 2(a). Away from topography a depth-uniform, nearly horizontal velocity is observed, as expected for barotropic tidal flow.
Amplification of the flow near seamounts is again visible over the west ridge. The acoustic surface probes revealed a phase difference of $\sim155^o$ upon passing across the strait, which is in good agreement with the $\sim180^o$ phase shift found in TPXO models and numerical simulations \citep{bib:Niwa2004}.
\begin{figure}
\centering
\includegraphics[width=0.95\linewidth]{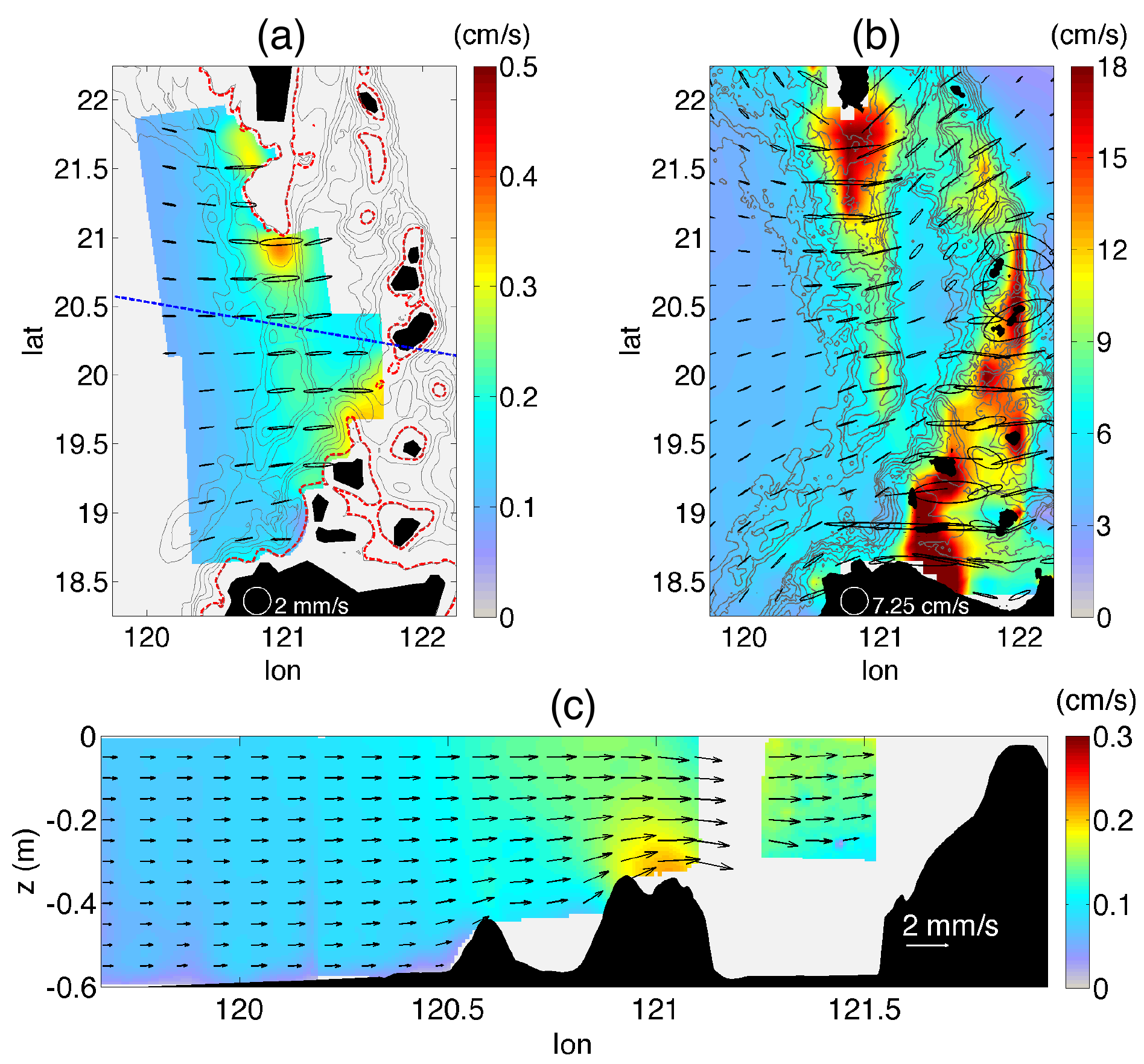}
\caption{M$_2$ tidal ellipses and horizontal velocity magnitude (a) in the experiment and (b) from TPXO 7.1. The ellipses and colorbars are scaled using the relevant scaling parameters listed in Table \ref{tab1}. (c) Vertical view of the velocity (arrows) and its time-averaged horizontal component (colormap) extracted along the W-NW transect (blue dashed line) in (a), when the tidal flow from the SCS to the PO is maximum.}\label{fig2}
\end{figure}
Finally, the introduction of a density stratification induced no discernible change in the depth-averaged velocities taken along the blue transect in Figure \ref{fig2}(a). Overall, we conclude that the barotropic tidal flow responsible for internal tide generation in the laboratory experiments successfully reproduced that in the ocean.

\section{The M$_2$ internal tide}
\label{sec:BC}
Figure 3(a) presents a snapshot of the east-west velocity in an isopycnal plane around $z=-0.04$m, with instantaneous velocity vectors inside the M$_2$ filtered tidal ellipses. This view essentially contains the baroclinic velocity component by virtue of considering the time instant of flow reversal of the barotropic tide (i.e. the onset of flow from the SCS to the PO). Large-scale coherent waves radiate from the strait in a west-northwest direction. The maps of the amplitude and phase of the M$_2$-filtered baroclinic and barotropic tides, which cannot be readily discriminated, are presented in Fig.\,3(b) and (c), respectively. There is a clear modulation of the amplitude oriented in the $\sim285^o$N direction, which is not due to interference with internal waves reflected at the wall of the tank, as no signal was observed in the SCS when filtering solely eastward propagating waves \citep{bib:Mercieretal08}. The apparent wavelength of the signal, 1.3$\pm$0.15\,m, is consistent with an internal tide dominated by the mode-1 wavelength of 1.4\,m for this stratification.
A notable feature in Fig.\,3(c) is that a wave front, evident as a sharp line of constant phase, connects coherently from the central and southern sections of the eastern ridge to the far field west  of both ridges, on the way smoothly adjoining the phase of the wave field generated by the northern and central sections of the western ridge; the combined wavefront coming from these sections of the two ridges propagates in a $\sim285^o$N direction. In the south, another distinct wavefront is oriented in a southwestward direction. In the northern section of the strait, between the two ridges, no clear phase propagation exists, a behavior characteristic of a standing wave pattern.
In the absence of background rotation, although the wavelength of the radiated internal tide is noticeably diminished, we found little discernible change in the direction of radiation, revealing the dominant role of the topography in shaping the radiated M$_2$ internal tide.

\begin{figure}
\centering
\includegraphics[width=0.95\linewidth]{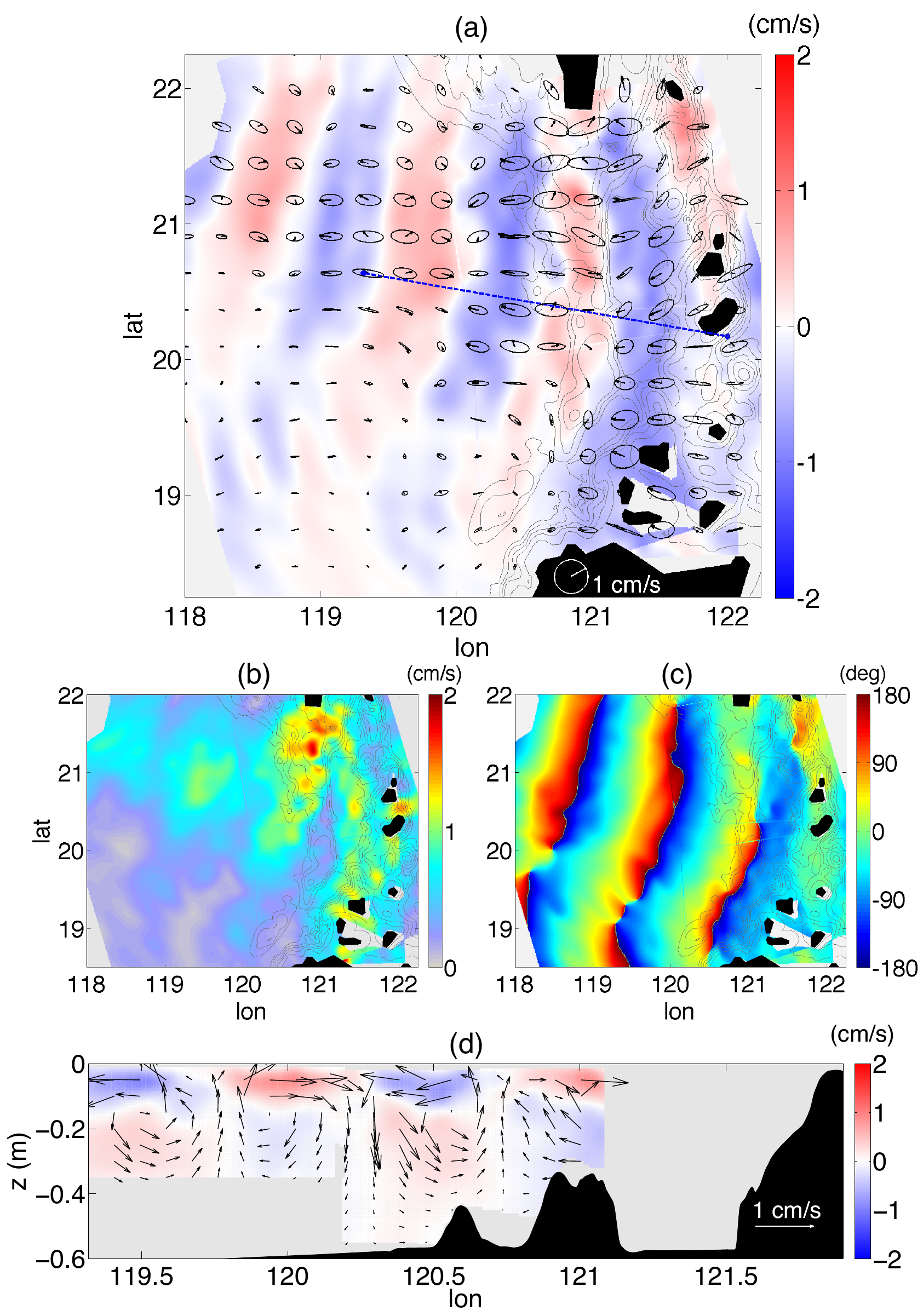}
\caption{(a) Colormap of the east-west velocity in the isopycnal plane at $z=-0.04$m at an instant of barotropic tide flow reversal. The local velocity direction is indicated by the black arrows inside the tidal ellipses. (b) Amplitude of the total velocity and (c) phase of the east-west velocity of the combined M$_2$ baroclinic and barotropic tides, filtered at the forcing frequency. (d) Same data as in (a) for the vertical transect indicated by the dashed blue line in (a), arrows indicate the in-plane velocity field.}\label{fig3}
\end{figure}

Figure \ref{fig3}(d) presents the same data as in Fig.\,\ref{fig3}(a), but for a vertical transect indicated by the blue dashed line. The structure of the velocity field takes the form of a classic mode-1 signal, with a vertical velocity maxima of alternating signs along the length of the transect and in the vicinity of the pycnocline, as expected from the observations in Fig.\,\ref{fig3}(a). Modal decomposition confirms these results, determining that $\sim95\%$ of the energy flux is in mode-1, and the remaining $5\%$ in modes 2 and 3 mainly.


\section{Nonlinearity}
\begin{figure}
\includegraphics[width=0.9\linewidth]{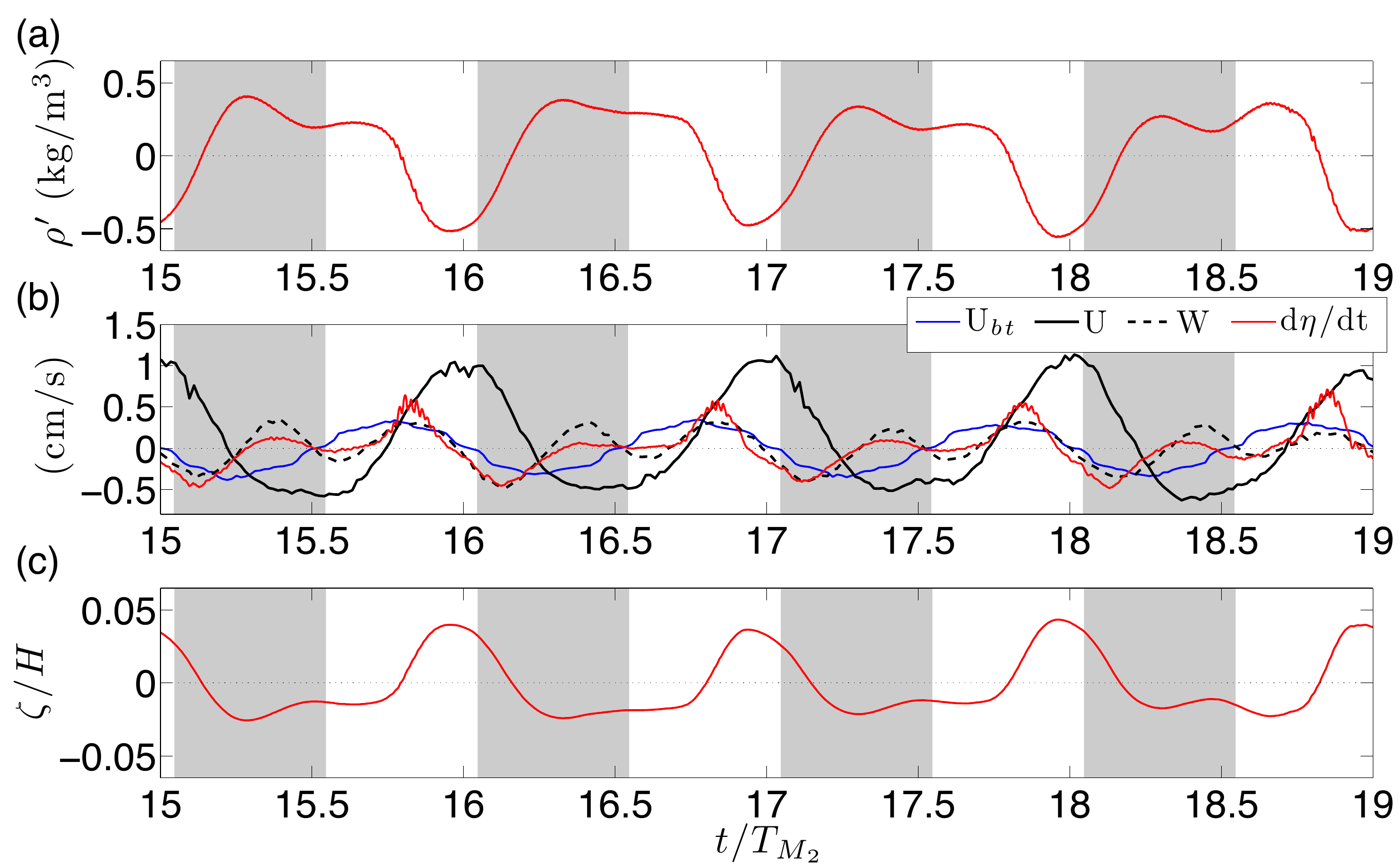}
\caption{Time series of (a) density perturbations $\rho^\prime$ from a conductivity probe located at $z=-3.6$cm and coordinates (20.5$^\circ$N,120.5$^\circ$E); (b) the east-west and vertical velocity (U and W respectively) at the same location, along with the barotropic east-west velocity U$_{bt}$ (experiments with no stratification) and the vertical velocity $d\eta/dt$ estimated from the vertical displacements $\eta$; and (c) the nonlinear first internal mode amplitude $\zeta$ compared to the water depth $H$. Gray (resp. white) regions correspond to flow from PO to SCS (resp. SCS to PO).}\label{fig4}
\end{figure}

Although the ISWs in the SCS are best described by fully nonlinear models \citep{bib:Li2011}, in order to reasonably quantify nonlinearity we assume the waves to be governed by a weakly nonlinear,  KdV-like equation.
The nonlinearity of the wave field is assessed by the ratio $\zeta/H$ (it is appropriate to scale with $H$ since the waves are dominantly mode-1), where $\zeta$ is a measure of the wave amplitude given by $\eta(x,z,t)=\zeta(x,t)\phi(z)$, $\eta$ being the vertical displacement of an isopycnal initially located at $(x,z)$, and $\phi$ the long wave vertical structure function of mode-1 evaluated at the initial depth of the isopycnal \citep{bib:Grimshaw1998}.
Estimations of $\zeta/H$ are based on time series presented in Fig.\,\ref{fig4}, extracted at $z=-3.6$ cm and coordinates (20.5$^\circ$N,120.5$^\circ$E).
Given the stratification profile, $\eta$ is obtained from the CT probe recordings of the density perturbation, $\rho^\prime$ in Fig.\,\ref{fig4}(a).
The vertical velocity estimated from the vertical displacement ($d\eta/dt$) is compared with the vertical velocity (W) obtained from PIV measurements in Fig.\,\ref{fig4}(b), showing a good correlation between the two signals, although stronger nonlinear oscillations are observed in $d\eta/dt$. The east-west barotropic and total velocities, U$_{bt}$ and U respectively, are also displayed to indicate the linear nature of the barotropic tide and the dominant baroclinic contribution in the total velocity.
Finally, the time series of $\zeta/H$ in Fig.\,\ref{fig4}(c) shows that the wave amplitude is $\sim5\%$ of the water height. This is consistent with the assumption of weak non-linearity and in good agreement with oceanic observations and a recent analysis of  \cite{bib:Li2011} based upon the rotationally-modified KdV equation.

The steepening of the radiated internal tide, producing sharp ISWs, is not observed in the experiments.
This is not due to a stronger dissipation, but rather the consequence of the modified aspect ratio in the experiments. The characteristic wavelength of a soliton with amplitude $\zeta_0$ is $L_S\sim\!(H^3/\zeta_0)^{1/2}$, while the wavelength of the linear mode-1 
is $\lambda_1\simeq \!H/\tan\theta$, with $\tan\theta=\!\sqrt{(\omega^2-f^2)/(N^2-\omega^2)}$. For our experiments ($\zeta_0/H = 0.05$), the ratio $L_S/\lambda_1 =  (H/\zeta_0)^{1/2} \tan\theta \sim 1$, meaning that ISWs have the same horizontal scale as mode-1, while in the ocean the horizontal scale of ISWs is roughly $1/25$ that of the mode-1 tide.
The experiments do not preserve the scale separation between ISWs and the internal tide required to observe the subsequent steepening.

\section{Conclusions and Discussion}

These experiments provide a clear demonstration that despite having complex three-dimensional geometry, the LS radiates a spatially coherent, mode-1 dominated, weakly nonlinear M$_2$ internal tide oriented in a west-northwest direction. The spacing of the ridges being on the order of the mode-1 internal wavelength for M$_2$, extended sections of both ridges shape the westward propagating M$_2$ internal tide.
We observed modulation of the amplitude of the radiated wave field due to interference between the barotropic and baroclinic tide, an effect observed
  in numerical models ~\citep{bib:Buijsman2010a}. The commensurate spacing of the ridges in the northern section of the LS results in a standing wave pattern.
Being weakly nonlinear, the radiated internal tide is prone to steepening, and this is the origin of the ISWs in the SCS, although the steepening process itself is not captured in these experiments.
These results do not preclude the ability of local topographic features to produce ISWs, observed on a smaller scale \citep{bib:Ramp2012}. We expect that further experimental work incorporating both the diurnal and semi-diurnal tides will provide insight on ISWs with differing characteristics, such as the reported A and B waves \citep{bib:Ramp2004}. Other potentially important considerations, such as the Kuroshio, could not reasonably be incorporated into the experiments.


\begin{acknowledgments}
We thank the IWISE collaboration for many productive interactions. This work is funded by ONR grants N00014-09-1-0282 \& N00014-09-1-0227, CNRS-PICS grant 5860, ANR grant 08-BLAN-0113-01, and the MIT-France Program.
\end{acknowledgments}

\end{article}


%
%

%
%
%
%
%
%
%


\end{document}